\documentstyle[psfig]{kluwer}
\def\verbfont{\small\tt}
\newcommand{\Rsun} {$R_{\odot}$}
\newcommand{\Msun} {$M_{\odot}$}
\newcommand{\Lsun} {$L_{\odot}$}
\newcommand{\Teff} {$T_{\rm eff}$}

\begin{document}
\begin{frontmatter}
\end{frontmatter}
\begin{article}
\begin{opening}

\title{P Cygni: An Extraordinary Luminous Blue Variable}
\author{G. \surname{Israelian}$^{1}$ and
M.\surname{de Groot}$^{2}$}
\institute{$^1$Insituto de Astrofisica de Canarias, E-38200 La Laguna, 
Tenerife,
 Canary Islands, Spain,\email{gil@iac.es}}
\institute{$^2$Armagh Observatory, College Hill, Armagh, BT61 9DG, Northern
Ireland, UK,\email{mdg@star.arm.ac.uk}}

\received{ 17 May, 1999}

\runningtitle{P Cygni: An Extraordinary LBV}
\runningauthor{G.Israelian and M. de Groot}

\begin{abstract}
{
P Cygni is a prototype for understanding
mass loss from massive stars.
This {\it textbook} star is known first of all because of
two great eruptions in the 17th century. In the first half of this
century it has given its name to a class of stars
which are characterized by spectral lines consisting of
nearly undisplaced emissions accompanied by a blue-displaced
absorption component. This characteristic P Cygni-type profile betrays the  
presence of a stellar wind, but P Cygni's wind is quite
unlike that of other hot supergiants. P Cygni was the first star 
that showed  the
effects of stellar evoluton from a study of its photometric history.
It shares some common properties with the so-called Luminous Blue Variables.
However, P Cygni is a unique object. 

This review deals with P Cygni's photometric properties, its circumstellar
environment - including infrared and radio observations - and its optical and  
ultraviolet spectrum. Smaller sections deal with P Cygni's wind structure and
evolution.
}
\end{abstract}

\end{opening}

\section{Historical overview}

About 400 years have elapsed since the Dutch chartmaker,
mathematician and geographer Willem Janszoon Blaeu  
recorded P Cygni (34 Cyg, HR 7763, HD 193237) 
as a nova. It was discovered on August
18$^{th}$ of the year 1600 when it suddenly reached 
3rd magnitude. After remaining at its maximum  
for about 6 years, the brightness started to decrease and
the star became invisible to the naked eye in 1626. 
It reappeared in 1654 and remained at 3rd magnitude
until 1659. The visual magnitude of the star was variable
between 1660 and 1683 and started to increase slowly
until 5.2 was reached in 1780. Except for a few observations by
Herschel in the 1780s, there is no record of variations 
between 1780 and 1870. Since then P Cygni's brightness
has increased slowly by 0.4 magnitudes to bring it to its current
value of V=4.83. Analysis of historical observations of P Cygni has shown 
that between 1700 and 1988 its overall brightness slowly increased  
by 0.15$\pm$0.02 magn/century (de Groot and Lamers 1992).
The light curve of P Cygni after 1600 AD is displayed in Fig. 1.

The first spectra of P Cygni, obtained as early as 1897, already
showed the famous P Cygni-type spectral lines (Fig. 2); an undisplaced emission
accompanied by a shortward displaced absorption core (Maury 1897). 
Initially this
was interpreted as a blend of two different lines. 
McCrea (1929) and Beals (1930a, 1932) were the first to
interpret P Cygni-type profiles in novae, Wolf-Rayet stars 
and P Cygni itself as due to a radially expanding stellar envelope.
Later investigations by Struve (1935), Beals (1935) and Struve and
Roach (1939) have shown that the outflowing matter is accelerated
throughout the observed region since the lower-excitation
lines, which presumably form in the remotest part of the envelope,
systematically have the highest observed outflow velocities.
Early serious detailed analyses of P Cygni's spectrum have been carried
out by Beals (1950), Hutchings (1969) and de Groot (1969).
Some identifications and measured equivalent widths from 
moderate-dispersion spectra were presented by Luud (1967a).
More recently, Johnson et al. (1978), Stahl et al. (1993) 
and Markova (1994) have published spectral atlases with
identifications of many weak lines in the visual spectral 
region. More than one hundred photographic spectra have been
obtained between 1980 and 1990 by N. Markova with the 
coud\'{e} spectrograph of the 2m RCC telescope at the Bulgarian 
National Astronomical Observatory. Another monitoring
programme has been carried out by Stahl et al. (1993, 1994) using
a fiber-linked echelle spectrograph with EEV CCD coupled 
to the Heidelberg (Germany), 
70 cm telescope in Tautenburg (Germany) and the 2.2 m 
telescope in Calar Alto (Spain). The study of high-resolution
and high-signal-to-noise (S/N) spectra has resulted in
an abundance of information about the radial-velocity
and line-profile variations of the spectral lines.

The {\it Copernicus} satellite obtained the first spectra of P Cygni
in the non-visible ultraviolet region (Hutchings 1976; Ambartsumian et al. 
1979).
A large number of high- and low-resolution ultraviolet spectra 
has been obtained with the {\it IUE} satellite and analysed 
by Cassatella et al. (1979), Hutchings (1979), 
Underhill (1979, 1982), Luud and Sapar (1980), Lamers et al. (1983), 
Lamers et al. (1985) and Israelian et al. (1996).

Near-IR and IR observations have been reported by 
Barlow and Cohen (1977), Abbott et al. (1984), Waters and Wesselius (1986), 
and Lamers et al. (1996). Barlow and Cohen (1977) used
infrared and radio observations in order to estimate 
the mass-loss rate. The radio data, which directly 
measure the amount of ionized gas beyond the acceleration
zone of the stellar wind, are able to provide a
firm estimate of the mass-loss rate so long as the 
terminal velocity of the wind is known. Submillimeter
and radio data have been obtained by White and Becker (1982),
Becker and White (1985) and Altenhoff et al. (1994) while 
Wendker (1987) has compiled all observations in the range 
0.33--20 cm. 

P Cygni is located in the upper part of the Hertzprung-Russell 
(HR) diagram populated by different types of emission-line
stars, including Of supergiants, O3If/WN6, Ofpe/WN9, B[e],
Luminous Blue Variables (LBVs) and Wolf-Rayet (WR) stars (Fig. 3).
It is clear that all these classes represent different phases
in the evolution of stars with zero-age main-sequence (ZAMS) masses
of more than 40\Msun. This is the least understood
phase of the evolution of massive stars. Though it is commonly
accepted that massive O stars ultimately evolve into
WR stars (Maeder and Meynet 1987), the details of the evolutionary 
connections in the intermediate phases are not yet understood. Objects like
P Cygni, $\eta$ Car, AG Car and S Dor are ideal laboratories 
to be used to understand all missing links in the evolutionary chain 
O $\rightarrow$ WR.    
A complete review of LBVs has been given by Humphreys and Davidson (1994). 
Note that the class of LBVs is not defined 
by some spectroscopic classification criterion and that, obviously, it
may contain stars in different evolutionary states. The fact
that several LBVs are located near the so-called 
Humphreys-Davidson (HD) upper luminosity limit on the
HR diagram (Fig. 3) suggests that 
very massive hot stars do not evolve into red supergiants but 
somehow reverse their tracks near the HD upper luminosity
limit after a significant fraction of their mass is lost.
The physics of the mass loss from LBVs is not yet 
understood. Almost all LBVs, except P Cygni, show moderate-amplitude
cyclic spectrophotometric variability with timescales
of years to decades. The amplitude of the photometric 
variability can reach up to 2 mag and stars can change their
spectral type from early B to F. It is not yet clear
what is happening with a star when it is changing its spectral
type. There are some reasons to believe that the bolometric luminosity
of LBVs may not be constant between minimum and maximum state.
Most curiously, some LBVs do not show a general correlation 
between mass loss and radius when they move on the HR diagram 
(Leitherer 1997). However, all these conclusions must be considered
as provisonal since the number of LBVs studied in detail during  
their moderate- or large-amplitide excursions on the H-R diagram is 
very small.            

Though P Cygni and $\eta$ Car are the first LBVs known, later discoveries of
other LBVs, e.g. S Dor and AG Car, has shown that the former are not really
representative of the class. While both should be classified as LBVs - massive
stars ($M_{\odot} \geq 25M_{\odot}$) in a critical phase of their post 
main-sequence
evolution when they become unstable and may eject a significant fraction of
their mass in a relatively short time - they do not share many properties.
The properties of $\eta$ Car have been the subject of 
a recent review paper by Davidson and Humphreys (1997). We feel it
necessary to restore some balance by presenting another review paper
which summarises our knowledge about another enigmatic object: P\~ Cygni.

\section{Photometric observations}

\subsection{The giant eruptions and long-term variability}

In their review paper, Humphreys and Davidson (1994) have
used the term {\it astronomical geyser} to describe highly
unstable processes going on in the atmospheres of
$\eta$ Car, P Cygni, S Dor, AG Car and other LBVs. 
Indeed, there is a similarity with volcanic geysers: 
moderate activity followed by violent burbling, then
the huge eruption followed by relative quiescence.
Only four LBVs, specifically $\eta$ Car, P Cygni, 
V12 in NGC 2403, SN 1961V, have had {\it giant} eruptions
in which the star brightened by more than two magnitudes.
More than one solar mass can be ejected during these
giant eruptions which may last for decades. 
The fact that 
only four LBVs have passed through the giant-eruption
phase can be interpreted in different ways. It is possible
that the time interval between giant eruptions is very long
(centuries) and that not all LBVs have yet experienced them. 
Another possibility is that not all LBVs show giant eruptions. 
However, AG Car, HR Car and R127 may have
had giant eruptions as is evidenced by the presence 
of massive and extended nebulosities around them
(Humphreys and Davidson 1994; Nota et al. 1995). 
Using the number of known LBVs and the number of known 
eruptions one can estimate a time-scale of giant 
eruptions $\sim$ 10$^{3}$ years. 

It is interesting to compare P Cygni's and $\eta$ Car's environments. 
Both stars have experienced dramatic outbursts a
few hundred years ago. The Homunculus of $\eta$ Car, formed by 
ejecta of the giant eruption 
that was observed between 1830 and 1860, is one of    
the most colourful bipolar nebulae known. 
P Cygni's nebula, on the other hand, is spherically symmetric and 
hardly observable at optical wavelengths (Leitherer and Zickgraf 1987;
Johnson et al. 1992). Recently, it has been resolved by imaging
 (Barlow et al. 1994; Nota et al. 1995; Meaburn et al. 1996). 
The question is: why do these two LBVs, $\eta$ Car and
P Cygni, have such very different nebulae?  Could it be that their 
giant eruptions are governed by different physics? 
No other LBV is similar either to P Cygni or to $\eta$ Car. 
The only reasons which make us place these two objects into 
the same evolutionary group are their locations on the HR diagram 
and the giant eruptions they have experienced.
Is this similarity enough to claim convincingly that
both stars are in the same evolutionary state?  

Our understanding of the physics of giant eruptions is
based primarily on historical observations. In the years after the 
17th-century eruptions, and aided by the increased use of telescopes at that 
time, many observers provided estimates of P Cygni's brightness. 
Of course, for the first three centuries after AD 1600 these were
visual estimates made with or without a telescope and by many different
observers. As a result, these early
photometric data are not very homogeneous. However, quite an extensive 
record of these observations has been assembled by the authors of the 
various issues of the ``Geschichte und Literatur des Lichtwechsels 
der veraenderlichen Sterne'' (M\"{u}ller and Hartwig 1918; Prager 1936; 
Schneller 1957). A number of additional observations between 
1596 and 1917 is given by Zinner (1926). From all these records a 
fairly accurate picture of P Cygni's secular photometric history can
be reconstructed.

The brightness of the star went through deep minima after the two 
outbursts in 1600 and 1655 AD (Fig. 1). The matter ejected in the major 
outbursts most likely condensed into dust rings which expanded into
space. The reddish colour of P Cygni during the
outbursts has been recorded by several observers. However,
various uncertainties do not allow us to distinguish
between a circumstellar and an intrinsic
origin of this colour. In other words,  
the historical observations do not provide enough
evidence to check whether the star had the same 
M$_{\rm bol}$ as at present or whether it was much
brighter. Two possibilities have been suggested
by Lamers and de Groot (1992). If the mass loss was high
during the outburst in 1600 AD, then
the reddish colour observed at that time could have been due
to the large circumstellar reddening and the larger value 
(at least one magnitude) of the bolometric
magnitude. However, if the circumstellar extinction was small
during the outbursts, the observed colour would indicate an
effective temperature as low as 5500 K, if
the bolometric magnitude of the star was the same as the
present value. It is very important to answer
the following question:
can a LBV like P Cygni evolve down to $T_{\rm eff}$ = 5500 K 
or must it stop near 8000--10\,000 K as many other LBVs do 
during their excursions on the HR diagram?
 
It is believed that LBVs at their maximum do not pass
beyond the opaque-wind limit (Davidson 1987) because the
opacity in the envelope decreases drastically below 
T$_{\rm eff} \approx$ 8000 K due to hydrogen recombination.
Langer et al. (1994) suggested that LBVs never
become cooler than about 10\,000 K because of the large
mass loss near the instability limit.

It is not yet clear which physical processes are responsible for 
giant eruptions in LBVs and we cannot rule out the possibility
that different mechanisms work in different stars depending
on rotation, luminosity, etc.  
The giant eruption of $\eta$ Car took place between 1830 
and 1860 when the star increased its total luminosity reaching
M$_{\rm bol} \approx -14$ . It remained of the first magnitude
for 20 years and then faded down to 8th magnitude. As from 1935 it 
has been brightening to its current value V$\sim$ 5.2 mag (van Genderen et
al. 1994; IAUC 7146). The 
luminous energy released during the Great Eruption ($\sim$ 10$^{49}$ ergs)
was about the same as in a supernova explosion. This eruption
was more dramatic then the one observed in P Cygni. Furthermore, the 
eruptions of P Cygni and $\eta$ Car have produced different types 
of nebulae (Nota et al. 1995). The fact that $\eta$ Car is more 
luminous than P Cygni does not seem to provide the full answer.
More likely it has something to do with
the geometry (binarity, magnetic fields, interacting winds?). 
Therefore, to the question raised by Davidson and Humphreys (1997) 
``{\it Is Eta really an LBV?}'' we can add another one:
``{\it Is P Cygni really an LBV?}''. 
Indeed, according to Nota et al. (1995) $\eta$ Car's bipolar 
nebula shares the morphological class of {\it Shell nebulae}
with many other LBVs while P Cygni is the only object in the 
class of {\it Peculiar morphologies}. So, in this respect,
P Cygni is even more peculiar than $\eta$ Car. This fact  
suggests that different physical processes would have caused
the giant eruptions of these two LBVs. 

It is believed that some sort of pulsational instability is driving 
an extra mass loss that later appears as a giant 
eruption. Recently, Langer et al. (1994) have discovered a
phase of violent radial pulsations when a very massive 
($> 40 M_{\odot}$) shell-H burning  star evolves away from the 
main-sequence and its effective temperature drops below
20\,000 K. Their calculations show that a star with 
M$_{\rm ZAMS}$ = 60 $M_{\odot}$ may lose 
5$\times10^{-3}M_{\odot}$ yr$^{-1}$ near $T_{\rm eff} \approx
20\,000$ K and will evolve again to hotter temperatures when about 
6$M_{\odot}$ has been lost. However, it is not clear 
whether such pulsations occur in stars with enhanced mass loss
(either giant or moderate eruptions) and  
$T_{\rm eff} \approx 8000 - 10\,000$ K (e.g. LBVs).  

It is known that all normal LBVs, i.e. with the exception of
P Cyg and $\eta$ Car whose photometric variations are of
much smaller amplitude, show moderate
photometric variability with $\Delta V \sim$ 1--2 mag. Recently,
Van Genderen et al. (1997) have introduced the concept of ``S Dor phases''
(SD), phases of brightening with an almost regular 
pattern of recurrence. They suggested a new nomenclature discerning
between ``normal S Dor phases'' superimposed on a much slower 
gradient of brightening and fading for which they introduced
the term ``very-long term S Dor phase'' (VLT-SD). The presence in
LBVs of two distinct oscillation time-scales may have a strong impact 
upon studies of pulsation,
stability and evolution of stars near the HD limit. 

An important step towards our understanding of the evolution of
massive stars was the discovery of a steady increase 
in P Cygni's visual brightness between 1700 and 1988
(de Groot and Lamers 1992; Lamers and de Groot 1992). Evaluation
of P Cygni's brightness during the period 1700-2000 is
a difficult task because the historical observations are based
on a variety of early photometric detectors and different
filters. Assuming a constant luminosity during 
the last three centuries, these authors have shown
that the increase in brightness of 0.15$\pm$0.02 mag/century
is caused by a steady decrease of the Bolometric Correction in
agreement with calculations based on the theory of stellar evolution.
This means that P Cygni has been moving to the red on the 
HR diagram changing its $T_{\rm eff}$ at a rate of 6$\pm$1
percent per century. Clearly, P Cygni is the first massive star
whose long-term, steady photometric changes have been shown to
be caused by evolutionary effects.

\subsection{Microphotometric short-term variability}

As for P Cygni's photometric behaviour on shorter time-scales, 
reports in the older (i.e. pre-1930) literature are almost unanimous 
in saying that the star is
essentially non-variable: Safarik (1888), e.g., reports P Cygni as ``just about
constant'' on 17 nights between July 1884 and June 1886, and without 
recognizable
variations thereafter until November 1888. Reports by other observers
give essentially the same verdict (e.g., Flammarion 1882; Gore 1884; 
Markwick 1892; Campbell 1940). As a result, there 
has been no encouragement to observe the star regularly and,
from what follows, it must be concluded that during the years before the mid
1930s we have most likely failed to observe certain episodes of more distinct
brightness variations.  

An early attempt at investigating P Cygni's short-term variations seems to have
been made by von Prittwitz (1900) who used a Z\"{o}llner photometer and,
despite finding seemingly irregular variations with an amplitude of up to
$0.^{m}4$ between 1898 and 1900, concluded that the star showed ``no 
appreciable
variations''. Later, however, after further observations between 1902 and 1907,
she concluded that ``P Cygni shows a tendency to very gradually become somewhat
brighter'' (von Prittwitz 1907). Von Pritwitz's gradient over her 9-year
observation period amounts to 2.1 magnitude per century, more than an order of
magnitude larger than the above-quoted figure found by de Groot and Lamers over
the much longer interval of almost three centuries.

Another notable exception to the generally indifferent attitude to P Cygni's
brightness is found in the work of the visual observers of the AAVSO. P Cygni
was included in  their observing programme as early as 1917 and
has remained there ever since with a respectable number of photoelectric
measurements added since 1985 (Percy et al. 1988). The visual estimates made by
the AAVSO observers suffer from some drawbacks, though. They were collected by 
a
large number of individual visual obervers mostly with insufficiently
known personal equations, making it difficult to construct a homogeneous light 
curve that accurately reflects the rather small-amplitude variations 
of the star.

The early years of the AAVSO observations were also the early years of
photoelectric photometry. However, in those early years there was no widely
accepted standard photometric system. Photoelectric detectors had very
individual characteristics and the choice of filters seems to have been
determined largely by whatever piece of coloured glass an observer could lay 
his
hands on. This, together with the absence of serious encouragement for the
photometry of P Cygni, must have been largely responsible for the lack of
photoelectric data that could in principle be tied in to a modern standard
photometric system like, e.g., the Johnston UBV system.

The situation improved considerably when the Abastumani observers started
systematic two-colour observations of P Cygni in the mid 1930s (Kharadze 1936;
Nikonov 1936, 1937). Though not all observations obtained since these earlier
ones have been published (Kharadze et al. 1952) and despite the fact that 
their
pre-1951 observations were done with non-standard filters, these observers
concluded that P Cygni must be a W Uma system with a period 
of $\sim$0.5 d
(Magalashvili and Kharadze 1967a,b). They ascribed the absence of any earlier
detection of this characteristic, and the belief that P Cygni's light 
variations
were wholly irregular, to the insufficient number of observations that had been
made thus far. However, as pointed out by Fernie (1968), the W\~UMa
interpretation is untenable because it requires a primary star with a mass of 
50
$M_{\odot}$ and a secondary with an orbital velocity of 10,000 km s$^{-1}$. 
Alexander and Wallerstein (1967) and Luud (1969) have provided further
photometric evidence and de Groot (1969) gave spectroscopic reasons why P Cygni
cannot be considered a simple multiple object.

Although the Abastumani observers seem to have arrived at a wrong conclusion,
one fact was established beyond reasonable doubt: P Cygni is a photometric
variable. Here it is good to remember that Dreyer, on the basis of a discussion
of William Herschel's observations of 1792 and 1795, came to the conclusion 
that
P Cygni does show some, probably irregular, variation in brightness (Dreyer
1918). Zinner, too, found that P\~Cygni ``apparently shows very rapid, 
substantial
brightness variations'' (Zinner 1938). The Abastumani results did encourage
later observers to dedicate some of their time to P Cygni (e.g., Groeneveld 
1947)
but there was never a long-term observing campaign until the 1980s when first
Percy and his co-workers (Percy and Welch 1983; Percy et al. 1988)  and then de
Groot (de Groot 1990) began observing P Cygni on
both a more regular basis and a longer time-scale. From these observations a
number of characteristics of P Cygni's light variations have become clear. 

\begin{description}
\item[a.]Over the last 15 to 20 years (and most likely over most of the time
since its two 17th century outbursts) P Cygni's variations have had a 
maximum amplitude of about $0.^{m}2$ in $V$. 
\vspace{-1mm}
\item[b.]Most of the time the variations are slow, say $\Delta V \leq0.05$ magn
   month$^{-1}$ in the V band.  
\vspace{-1mm}
\item[c.]On a time-scale of months to a year there are larger, more rapid
   variations, say $\Delta V \sim 0.15$ magn month$^{-1}$ in the $V$ band. 
\vspace{-1mm}
\item[d.]The overall light-curve shows a number of 
quasi-periodicities ranging from 17 to $\sim 1500$ d (de Groot et al. 1999)
\end{description}

\section{Circumstellar matter}

Nebulae around LBVs are of considerable interest since they provide
clues to the mass-loss history of massive stars before they reach
the LBV phase.
P Cygni's nebula is very faint at optical wavelengths 
and composed of a distribution of "blobs" within a spherically symmetric 
nebula with a diameter of 22$''$.  Early detection of 
nebular emission lines [S\,{\sc ii}] 6716, 6731 \AA, [N\,{\sc ii}] 6584 \AA\
in a long slit spectrum offset 9 arcsec east from P Cygni was achieved
by Johnson et al. (1992). They found a ratio N/S of about 33$\pm$5
by number from the [N\,{\sc ii}] and [S\,{\sc ii}] line fluxes. An
anomalously strong nebular [Ni\,{\sc ii}] 6667 \AA\ line was detected by
Johnson et al. (1992) and by Barlow et al. (1994). Lucy (1995) has
explained the strength of the [Ni\,{\sc ii}] 7378 and 7412 \AA\
emission from the fact that the [Ni\,{\sc ii}] emission in ejected
globules is being fluorescently excited by the near-UV radiation
of P Cygni. According to Barlow et al. (1994) bow-shocks around
these globules, generated by the radiatively driven wind from the
star, are responsible for the difference between the [N\,{\sc ii}]
6584 \AA\ and [Ni\,{\sc ii}] 7378 and 7412 \AA\ expansion velocities of 
140 and 110~${\rm km}~{\rm s}^{-1}$, respectively. Two faint outer arcs
(Fig. 4) with angular diameters 1$'$ and 1.5$'$, respectively, have been 
discovered
(Barlow et al. 1994). Meaburn et al. (1996) have measured a radial 
expansion velocity of 185~${\rm km}~{\rm s}^{-1}$ of the outer
irregular shell located at 1.6 arcmin.
The presence of two inner rings detected through the [Ni\,{\sc ii}] filter 
and having radii of $\sim$ 11$''$ and 6$''$ has also been suggested
(Barlow et al. 1994). If confirmed, a $\sim$ 6$''$ ring could date back
to the 1600 AD event. The shell located at 22$''$ has probably been 
ejected 900 years ago. More recently, based on MOMI observations with 
the NOT (La Palma), O'Connor et al. (1998) found another filament at 
a distance $\geq$ 5$'$ from the star. This filament was investigated further
by Meaburn et al. (1999). They concluded that this filament, which extends 
only to the east of P Cygni, seems the relic of a highly asymmetric 
ejection some 20,000 years ago. However, the exact nature of this filament 
and how it was formed remains unclear, leaving several questions
unanswered, not the least of which is the fact that this early outburst 
seems to have happened much longer ago than one would expect P Cygni's 
LBV phase to have lasted.  Nevertheless, with shells from outbursts 
$\approx$ 900, $\approx$ 2100 and, maybe $\approx$ 20,000 years ago, a 
possible scenario for the episodic eruption history of P Cygni is now
taking shape.

Some LBVs exhibit an IR excess due to dust emission. At 60 $\mu$m
one would expect at least 7-8 Jy  from the P Cygni
nebula, assuming the same ratio of dust  to ionized gas as observed
in other galactic LBVs. However, Waters and Wesselius (1986) obtained
2.05 Jy corresponding to an excess of only $\sim$ 0.5 Jy above the
expected free-free emission from the stellar mantle at this wavelength.
The observed infrared energy distribution of P Cygni
does not show any signature of dust emission and must be
due to the free-free and bound-free continuum of the wind. 
This fact can be used to construct the velocity law and derive a
mass-loss rate. Using the ground-based IR fluxes (0.98--20$\mu$m)
by Abbott et al. (1984) and the {\it IRAS} fluxes (10--100$\mu$m),
Waters and Wesselius (1986) have concluded that the best fit to the   
observed infrared energy distribution could be obtained for
a linear velocity law. This result has been disputed by Pauldrach
and Puls (1990) and by Najarro et al. (1997) who showed that a typical
$\beta$-velocity law 
[$v(r) = v_{\rm s}+(v_{\infty}-v_{\rm s})(1-R_{\rm s}/r)^{\beta}$]
can reproduce the observations and also 
account for the wind acceleration. 

Recently, Lamers et al. (1996) have reported the {\it ISO}
observations of P Cygni covering the range 2.38--45.2 $\mu$m.
They found a good agreement comparing the observed profiles
of H and He lines and the free-free emission with the
predictions from non-LTE atmospheric models of 
Hillier (1987, 1990). Numerous forbidden lines of singly and
doubly-ionized Fe, Ni, Ne and Si have been identified.

The first radio observations of P Cygni, made by Wendker et al. (1973), have 
clearly revealed the source. The radio emission has been resolved by the 
VLA (White and Becker 1982) and these observations also confirmed its thermal
free-free nature. Wendker (1982) and Baars and Wendker (1987) 
have detected an arc of faint, thermal radio emission 30$''$ to the
north-west of P Cygni. This arc has not yet been identified at
optical wavelengths. It is well known that for a purely thermal 
free-free and bound-free emission model of the wind, the observed flux 
should follow a power law (e.g. Wright and Barlow 1975).  
Compilation of the observations by Altenhoff et al. (1994, 1.3mm)
White and Becker (1982, 6cm) and  Becker and White (1985, 2cm) done
by Wendker (1987) have been used by Pauldrach and Puls (1990) and
Najarro et al. (1997) to model the wind emission. These authors have
shown that the measured radio flux is predominantely of thermal nature
and can be used to measure the mass-loss rate.  The radio flux varies on 
short time-scales of the order of months (Abbott et al. 1981; 
van den Oord et al. 1985). The finding of Wright and Barlow (1975) 
that P Cygni's optically thick radio photosphere has a diameter $\sim$0.1$''$
at 6 cm has recently been confirmed by Skinner et al. (1997) who were 
able to resolve the radio core of the stellar wind at 0.07$''$ 
and show that it is very clumpy.

Polarimetry is another powerful tool to investigate
the circumstellar evironment of early-type stars.
Hayes (1985) was first to obtain broadband polarization
measurements of P Cyg in 1978 and 1979 and to discover an 
intrinsic polarization of the star. His results suggest
temporal variations of the amplitude of the polarization
and the scattering angle. Later, Taylor et al. (1991) 
reported the first optical linear spectropolarimetric 
measurements from 3200 \AA\ to 7800 \AA. They confirmed 
Hayes' (1985) earlier results and found large night-to-night 
variations in continuum polarization indicating that
the scattering sources lie close to the star. Both, Hayes (1985)
and Taylor et al. (1991a) have concluded that the circumstellar 
wind of P Cyg is inhomogeneous and that the shape of the
polarization versus wavelength curve varies with time.
First UV (1400-3200 \AA) spectropolarimetry of P Cygni 
(Taylor et al. 1991b)
was obtained with the Wisconsin Ultraviolet Photo-Polarimeter 
Experiment (WUPPE) on board of the ASTRO-1 space observatory. 
These observations show that the intrinsic polarization of the
star remains constant in the Balmer continuum except for a broad 
dip between 2600 and 3000 \AA. This result was not predicted by 
theoretical models (e.g. Cassinelli et al. 1987) and explained 
by Taylor et al. (1991b) in terms of Fe line blanketing 
between 2600-3000 \AA. The intrinstic polarization angle 
was also found to be constant in the UV except for a rotation
accross the feature between 2600-3000 \AA. Models to explain
polarization data obtained by WUPPE have been developed
by Bjorkman and Cassinelli (1993).

\section{Optical spectroscopy}

The spectrum of P Cygni shows undisplaced emission lines flanked by absorption
lines displaced to shorter wavelengths. This characteristic has been the reason
for the term `P Cygni profile' whenever a similar composite profile appears in
the spectrum of some cosmic source. The P Cygni characteristic is shown most
completely in P Cygni itself: almost all the lines in its optical spectrum show
a P Cygni profile. The spectrum is dominated by H, He\,{\sc i},
N\,{\sc ii}, N\,{\sc iii}, Si\,{\sc iii}, Si\,{\sc ii} and 
Fe\,{\sc iii} P Cygni profiles. Pure emission lines due to
[N\,{\sc ii}], N\,{\sc ii}, Fe\,{\sc ii}, Fe\,{\sc iii}, 
Mg\,{\sc ii} etc. have been identified as well. 
The earliest mention of this double structure goes back to
1897 (Maury 1897) when it was first interpreted as an overlapping of two
different spectra. Between 1888 and 1911 several observers studied the spectrum
of P Cygni. Merrill (1913a) did not include P Cygni in a list of stars with
variable spectra nor in a list of stars with spectra suspected of
variability. This parallels the early conclusions about P Cygni's photometric
behaviour. While different early observers had all come to this conclusion, we
should not forget that spectrogrammes of those days did not have a quality
comparable to what could be obtained in later years. Therefore, the early
conclusion about the non-variability of P Cygni's optical spectrum is not 
really
a proof that nothing has changed; but the variations must have been small.

In 1913, Merrill (1913b) was the first to report spectral variations after a
study of spectrogrammes obtained at Lick in 1912 and 1913. Frost (1912), on the
contrary, concluded that there were no variations on the basis of a study of a
number of spectrogrammes obtained between 1904 and 1912. Lockyer (1924) was of 
the
same opinion. Elvey (1928 and references therein), on the other hand, showed
that some variations in the spectrum seemed to have been observed. However, he
attributed these to instrumental errors. 

More recently, the question about spectral variations of P Cygni has been
answered positively and settled conclusively (e.g., Kupo 1955; Dolidze 1958; 
Luud 1967a,b; de Groot 1969; Stahl et al. 1994, and references in these). 
Other papers
from this period address P Cygni's absolute magnitude (Beals 1950: $M_{v}$=
$-$6.3; Arkhipova 1964: $-$7.7; Lamers et al. 1983: $-$8.3).

The interpretation of P Cygni's spectrum started in the 1930s with the works of
McCrea (1929) and
Beals (1930a,b, 1932, 1934, 1935), who explained the typical P Cygni profile as
due to a radially-expanding atmosphere around the star, and Struve (1935), who
presented the basic physical interpretation of the atmosphere. The recognition
that P Cygni's stellar wind is accelerated came when it was found that there is
a clear relationship between the excitation potential of a certain transition
and the radial velocities of its absorption lines (Beals 1935; Struve 1935;
Kharadze 1936). 

After some intermittent progress in the years following 
(Adams and Merrill 1957), progress accelerated 
beginning in the 1960s when a number of studies of the star's spectral
variations were published (Luud 1966, 1967a,b, 1969;
Markova 1993a). The first and most complete line identification 
in the range 3100 $<$ $\lambda$ $<$ 8750 \AA\ was made by
de Groot (1969). 
More recently, Markova and Zamanov (1995) and Markova (1994) published 
spectral atlases with line identifications in the ranges 4840 to 6760 \AA\
and 3550 to 4800 \AA, respectively. However, all these observations 
were made on photographic plates. Results of the long-term (1990-1992)
spectroscopic monitoring of P Cygni with a spectral resolution of
12,000 and S/N $>$ 100 in the wavelength range 4050--9050 \AA\ have
been presented by Stahl et al. (1993, 1994). By averaging more than 100
CCD spectra they achieved a S/N-ratio up to 2000. Such  high-quality
spectra have allowed a detection of a large number of [Fe\,{\sc ii}] lines
(Stahl et al. 1991; Israelian and de Groot 1992) formed in the outer
parts of the stellar wind. Most of these emission lines have a complex 
structure due to blends (Israelian 1995). Surprisingly, such a high
S/N-ratio of the combined spectrum has revealed very weak absorption 
lines of high-excitation O\,{\sc ii} on the top of the flat-topped 
[Fe\,{\sc ii}] lines. These absorption lines of oxygen 
form deep in the atmosphere and can be used to measure the abundance
of oxygen. Despite many attempts to identify all spectral lines
in the spectrum of P Cygni, many lines still remain unidentified
(Johnson et al. 1978; Israelian and Nikoghossian, 1993; Stahl et al. 1993).
The amplitudes of the spectral variations detected by Stahl et al. (1994)
are not exceeding $\pm$ 30\% for the emission-line fluxes and
$\pm$ 30 km~sec$^{-1}$ for the radial velocities. Nevertheless, 
some larger spectral variations must have occurred during the last
decades. For example, comparing different identification line lists,
Markova and de Groot (1997) have concluded
that more than 70$\%$ of the pure emission lines in P Cygni's spectrum
have appeared only recently. Changes in the degree of ionization  
in the wind can also be the cause of the variations of  
extended wings of H$\alpha$ (Israelian and de Groot 1991; Scuderi et al. 1994).
It is becoming clear that the spectral variations of P Cygni are not
limited to DACs and line-profile variations. 

\section{Ultraviolet observations}

One of the puzzling features of the UV spectrum 
is the lack of P Cygni profiles, whereas the visual spectrum
is full of them. However, some resonance and low-excitation 
lines of Mg\,{\sc ii}, Fe\,{\sc ii}, C\,{\sc ii} etc. do have 
P Cygni-type profiles. Ambartsumian et al. (1979) have argued that the 
lack of emission is probably not a result of very high density 
(required to collisionally de-excite the upper levels of the UV 
transitions) in the wind as proposed by Hutchings (1979). 
The maximum velocities corresponding to the blue end of the absorption 
component have velocities up to 300~${\rm km}~{\rm s}^{-1}$ 
(Cassatella et al. 1979; Underhill 1979; 
Ambartsumian et al. 1979; Hutchings 1976). 
The resonance lines of C\,{\sc iv} at 1548 and 1550 \AA\  and
those of Si\,{\sc iv} at 1398 and 1402 \AA\ show weak P Cyg profiles 
superimposed on the wide photospheric profiles with extended wings
due to blends (Israelian 1995). All stellar absorption lines 
in the spectrum are shifted to the blue. Singly-ionized metals 
have narrow (FWHM 60 ~${\rm km}~{\rm s}^{-1}$) profiles shifted 
to $-$220~${\rm km}~{\rm s}^{-1}$, whereas the doubly-ionized metals
have broad (FWHM 110 ~${\rm km}~{\rm s}^{-1}$) profiles displaced
by about $-$80~${\rm km}~{\rm s}^{-1}$. The fact that doubly-ionized 
metals have larger width and smaller blue shift indicates that they are 
formed in the inner part of the wind. This follows from the clear  
relationship between expansion velocity and excitation potential
indicating that the excitation temperature decreases outward
(Beals 1950; de Groot 1969; Cassatella et al. 1979; Underhill 1982).
Many Fe\,{\sc ii} lines have asymmetric absorption lines, deepest at
$-180 {\rm km}~{\rm s}^{-1}$ with a sharp cut-off towards the violet
and a gradual decrease towards the red at $0~{\rm km}~{\rm s}^{-1}$. 
An excitation temperature in the envelope 
of 12000$\pm$900 K has been derived from curve-of-growth analysis 
(Cassatella et al. 1979). There is no evidence of highly-ionized 
species such as N\,{\sc v} in the wind, unlike many other B1 supergiants
(de Jager 1980; Underhill 1982; Ambartsumian et al. 1979; Cassatella et al. 
1979).
The electron temperature in the wind is lower (T$_{e}$
= 12000 K) than the effective temperature of the star (T$_{\rm eff}$
= 19.000 K) suggesting a possibility that the wind is in radiative equilibrium 
with the photosphere (Ambartsumian et al. 1979). Thus, the degree of
ionization in P Cygni's wind is lower than in the photosphere.
In fact, the spectral type of P Cygni derived from the visual
spectrum is generally reported as B1Ie (Lesh 1968; de Groot 1969; 
Lamers et al. 1983; Underhill 1982). However, there are some
spectral indicators in the visual (Struve 1935) and ultraviolet 
(Underhill 1982) range suggesting that later types may be appropriate. 
Ultraviolet spectra have allowed the conclusion that the nature
of P Cygni's mass loss is different from that of other supergiants
of the same spectral type. The mass ejection may have a dynamical
origin whereas radiation pressure acts on the ejected material
when its velocity is large enough to shift the wavelengths of the 
strong lines away from those of their photospheric counterparts. According to
Abbott (1977), relatively low ionization in the wind increases the number
of strong lines driving a wind, once overlapping with the photospheric 
lines is overcome. P Cygni has a 7--8 times lower terminal
velocity and a 3--10 times higher mass-loss rate than other 
B1 supergiants (de Jager 1980; Underhill 1982). This apparently is the cause 
of the lower degree of ionization in the wind.

\section{Discrete Absorption Components (DACs)}

\subsection{Optical DACs}

Perhaps the most remarkable feature in the spectrum of P Cygni
is the existence of discrete absorption components (DACs) of 
H, He, C, N,
Fe and many other elements. DACs have a FWHM about 10--15
${\rm km}~{\rm s}^{-1}$ and, therefore, only high-resolution
observations can reveal them. The first extensive 
analysis of multiple absorption components was made by
de Groot (1969) who observed three variable componenets in the hydrogen
lines at $-215$, $-160$ and $-95~{\rm km}~{\rm s}^{-1}$ and two 
in the helium lines. He found that the first component at
$-215$~${\rm km}~{\rm s}^{-1}$ was showing periodic radial-velocity 
variations with P=114 days and semi-amplitude 30~${\rm km}~{\rm s}^{-1}$.
No periodic variations were found in other components.
Luud et al. (1975) have argued that the second component at 
$-130$~${\rm km}~{\rm s}^{-1}$ was also variable, but with 
P=57 days. They failed to detect any periodicity in the first component
at $-215$~${\rm km}~{\rm s}^{-1}$. According to de Groot (1969)
and Luud et al. (1975) the velocity of a given component increases and 
decreases with time. A different scenario was proposed by Kolka (1983),
Markova and Kolka (1984) and Markova (1986b). These authors 
proposed that the variations of DACs are due to the ejected 
shells with velocities increasing in time up to a constant value.
This idea was supported by Lamers et al. (1985) and 
Israelian et al. (1996) on the basis of high-resolution {\it IUE} spectra.
Markova (1986a) has reanalysed the velocities of DACs of the Balmer 
lines published by de Groot (1969) and Luud et al. (1975) and has
shown that the data seem more consistent with the recurrent 
shell-ejection scenario of Kolka (1983) over the period 1981 to 1983.
She derived an ejection time of 200 days from the study of DACs 
observed in Balmer lines. This period was confirmed by Kolka (1998). 
Van Gent and Lamers (1986) have reanalysed published
radial-velocity measurements of the shell components of 
P Cygni and found that the variations are not due to periodic oscillations
but to shell ejections with a mean time-scale of 
60--75 days. From the study of 13 helium lines over three years, 
Markova (1993b) has concluded that DACs appear every four or 
five months. Slow profile variations of helium lines have been 
detected as well. In Table 1 we list all available observations 
of DACs observed in hundreds of optical and at least 80 high-resolution 
{\it IUE} spectra. In another study, Markova and Kolka (1985) 
and Markova (1993a) concluded that virtually all absorption
lines of H, He\,{\sc i}, N\,{\sc ii}, O\,{\sc ii}, Si\,{\sc iii}
and Si\,{\sc iv} show complicated profile variations which can be 
explained in terms of fast-moving DACs superimposed on a 
slowly-varying underlying profile. These slow profile changes are most
probably related to the variations detected by Stahl et al. (1994)
and can be caused by non-radial pulsations. 

\subsection{DACs in the Ultraviolet}

There have been a number of detections (Cassatella et al. 1979; 
Luud and Sapar 1980) of DACs in the UV lines
of  Cr\,{\sc ii}, Ni\,{\sc ii}, Ni\,{\sc iii}, 
Si\,{\sc iii}, Mn\,{\sc ii} etc. (Table 1). However, detailed studies
have been performed only of those detected in Fe. 
Variable DACs in  Fe\,{\sc ii} and  Fe\,{\sc iii} have been studied
in 33 high-resolution {\it IUE} spectra by Lamers et al. (1985).
They found that  Fe\,{\sc ii} lines have two absorption components, 
a stable and a variable one, and explained this phenomenon in terms 
of variable thick shells ejected from the star with a frequency of 
about once per year. The stable component
at $-210~{\rm km}~{\rm s}^{-1}$ did not vary over 5 years. The second
DAC varied from $-$112 to $-174~{\rm km}~{\rm s}^{-1}$ in 1978/79.
They did not find a stable absorpion component in Fe\,{\sc iii} lines
and concluded that the latter form in variable shells. With hindsight, 
it is now clear that this failure was due to their use of blended iron 
lines in their analysis. When one uses all available iron lines, 
two components are clearly seen in the Fe\,{\sc iii} lines (see Fig. 5).
More recently,
Israelian et al. (1996) have analysed 49 high-resolution {\it IUE} 
spectra obtained in the period 1985--1991 and found a repetition
time between two successive shells of 200$\pm$11 days. This value
agrees perfectly with those obtained from Balmer lines by
Markova (1986a) and by Kolka (1998). It has also been shown that 
Lamers et al. (1985) used blended iron lines in their analysis 
and failed to find Fe\,{\sc iii} and Fe\,{\sc ii} lines with
two and three absorption components, respectively. A clear example
of the Fe\,{\sc iii} line with two components is presented in Fig. 5.
It has been shown (Israelian et al. 1996) that the acceleration of
DACs in the earliest phases when they appear at $\sim$ 
$-50~{\rm km}~{\rm s}^{-1}$ can be as large as 
0.6$\pm$0.3$~{\rm km}~{\rm s}^{-1}~{\rm d}^{-1}$, while it is
about 0.1$\pm$0.05$~{\rm km}~{\rm s}^{-1}~{\rm d}^{-1}$ at the end.
Four shells have been clearly identified on the basis of the
gradual variations of DACs in Fe\,{\sc ii} lines (Fig. 6).

\subsection{On the origin of DACs}

P Cygni is the only star in which DACs have been followed 
for more than 30 years. In spite of all observational studies
of DACs, the mechanism of their formation is still unknown. 
DACs can be produced by spherically symmetric shells, 
or asymmetric blobs, puffs, etc.
In this paper we will use the term {\em shell}, keeping in mind that
these can be blobs, etc. Shells are denser and have smaller 
accelerations than the mean wind material.
  
At least three mechanisms have appeared in the literature
aimed at explaining the shell-ejection phenomenon in P Cygni. 
Lamers et al. (1985) used the ISW model of Kwok et al.
(1978) assuming that shells gain their momentum from the 
fast wind accelerated between the shell
and the star. This mechanism neglects radiation pressure acting 
on the shells and suggests that shells will 
increase their mass in time (because of the interaction with the
fast-moving wind material) and will eventually reach a constant 
velocity. Another, simplified model was proposed by Kahn (1989) 
who estimated  that Lyman alpha radiation is able to account for the
acceleration of shells observed in the optical region. 
Kahn (1989) and Lamers et al. (1985)
did not make any suggestions on the shell--formation mechanism.
Finally, Pauldrach and Puls (1990) suggested a so called 
{\it bi-stability} mechanism in
which shells represent a high mass--loss
and small $v_{\infty}$ state of the wind which is optically
thick in the Lyman continuum. According to their models,
the high mass--loss state of the wind is caused by
the radiation pressure on singly-ionized metals. The interaction
between fast and slow material was neglected in the bi-stable  
models of Pauldrach and Puls (1990). They have demonstrated
that P Cygni's wind is very unstable with respect to small
changes in the luminosity ($> 3\%$) or radius ($> 5\%$). 
This finding has led them to propose a cyclic mechanism 
for the shell-ejection phenomenon. According
to this mechanism, the photospheric temperature of the
star is controlled by a moving shell due to the wind-blanketing
effect. The photospheric temperature increases during
a high mass--loss phase (e.g. by backscattering from a shell)
until the hydrogen becomes ionized. This leads
to a decrease of the line force since the Lyman continuum
is optically thin and there are not many lines from 
singly-ionized metals to drive the wind. Thus, the mass-loss rate 
switches back towards its lower value and the wind blanketing
decreases together with the photospheric temperature.
Now again, since the photospheric temperature is low, 
the hydrogen will recombine and the wind blanketing will
start to increase leading to the formation of a new shell.
This bi-stability mechanism suggested by Pauldrach and Puls (1990) 
is able to explain some of the observed characteristics of the 
shell--ejection phenomenon. In particular, 
according to the bi-stability model, the shell ejection must be 
accompanied by photometric changes of the star due to the 
variations of the surface temperature. Such a correlation has 
been suggested by Israelian et al. (1996).

Our knowledge of the geometry of the shells is very poor.
Direct imaging by Barlow et al. (1994) has resolved
dense blobs (knots) with masses of 10$^{-5}$\Msun\ and
radii of 7.42$\times$10$^{5}$\Rsun\ at 7$''$ from the star, moving 
with a velocity 100--140 ${\rm km}~{\rm s}^{-1}$. 
It is not yet clear whether these blobs can be identified with the 
DACs observed in UV and optical lines. It is interesting to
mention in this respect, that Vakili et al. (1997) using
the Grand Interf\'{e}rom\`{e}tre \`{a} 2 T\'{e}lescopes
(GI2T) found a bright localized blob of moving gas at
4 photospheric radii to the south of P Cygni. 

Similar DACs have been widely observed in the UV spectra of 
many other hot, massive stars (Howarth and Prinja 1989). 
However, in contrast with other OB-type supergiants, DACs in 
P Cygni's spectrum do not appear in the resonance lines of 
Si\,{\sc iv} and C\,{\sc iv} but in the subordinate lines 
of metals. Also, DACs in P Cygni can be followed for 
weeks or months while those detected in normal OB supergiants 
have time-scales of the order of hours to days. Finally, it seems that
in all instances the DACs observed in P Cygni were much broader than
those observed in other early-type supergiants. 

The intrinsic
instability of the winds provide the basis for numerical
simulations of this behaviour (Owocki 1998). The involvement 
of non-radial pulsations (NRPs) in this phenomenon may contribute
an injection of seeds of instability into the wind.
Several continuous {\it IUE} monitoring campaigns of a limited 
number of objects concluded that the variability of the wind is 
probably coupled to the rotational time-scales. In these cases
a cyclically repeating pattern in the development of DACs
was observed. However, their phase coherence within the
few cycles was relatively uncertain. The rotation periods
normally exceed the repetition time-scales of DACs by a small
factor and the uncertainties in both quantities do not
allow one to conclude whether their ratio is a whole number or not.
A whole number could have been due to co-rotating
surface magnetic structures while a non-integer ratio would
indicate the presence of NRPs. The problem is still open
and more multiwavelength observing compaigns are required
to resolve it. Returning to P Cygni, we note that there is
an excellent database for a theoretical study of DACs
in P Cygni as long as one can reduce all observations of
DACs (Table 1) to the same scale.

\section{Structure of the atmosphere and the wind}

The goal of any modelling of an expanding atmosphere is to 
investigate how line profiles and the emergent energy 
distribution behave as a function of mass loss, luminosity
and a velocity law. Many analyses, of varying degrees of 
approximation, have been carried out on the spectrum of P Cygni. 
Kuan and Kuhi (1975) attempted to fit broad H$\alpha$ profiles
adopting the velocity law from the study of Castor et al. (1975)
and concluded that P Cygni's wind has a deceleration zone.  
This result was difficult to reconcile with the observed
velocity-excitation correlations (Beals 1950) and with the 
infrared index (Wright and Barlow 1975; Barlow and Cohen 1977). 
Later, Van Blerkom (1978) and Kunacz and Van Blerkom (1979)
have succeeded in fitting computed Balmer-line profiles to those
observed by means of an accelerating flow. Nugis et al. (1979)
have proposed a {\it three-zone} model for P Cygni's wind.
According to their model, the material in the first zone
is accelerated to about 300~${\rm km}~{\rm s}^{-1}$ by some
pulsation instability, then 
decelerated to 200~${\rm km}~{\rm s}^{-1}$ in the second zone
due to gravity and then again accelerated to
1000~${\rm km}~{\rm s}^{-1}$ by radiation pressure.
More recent papers on this topic are those of Drew (1985),
Lamers (1986), Pauldrach and Puls (1990), Israelian et al. (1993), 
Scuderi et al. (1994) and Najarro et al. (1997). Some of these 
studies have given conflicting results and the situation will
not become clearer until more accurate data and more sophisticated
radiative-transfer codes have become available. 

The successes and shortcomings in this area have been outlined
in the PhD thesis of F. Najarro (1995) who utilized the
code of Hillier (1987, 1990) based on an 
iterative technique. This code solves the transfer equation
in the co-moving frame subject to statistical and radiative
equilibrium in an expanding, homogeneous  and
spherically-symmetric atmosphere. It is believed
that the main success towards modelling of P Cygni's 
emission-line spectrum is based on the validity of the
approximation of spherical geometry. Polarization observations
by Taylor et al. (1991) and Hayes (1985) suggest that there 
is no preferred axis of symmetry.  However, there are 
some difficulties typical for a star located at the position of
P Cygni on the HR diagram. As we have mentioned above, 
Pauldrach and Puls (1990) found
that small changes in the stellar parameters (e.g., a
change of only 0.05 dex in \.M\ or L$_{*}$) can lead to  
dramatic changes in the ionization structure of the wind. 
Najarro et al. (1997) have concluded that in models in which
hydrogen is recombined the Lyman lines remain optically thick 
and, thus, can account for the absorption dips on the Balmer
lines. They found that models where hydrogen is ionized throughout the
wind fail to fit the observed profiles of the H and He lines.
However, we want to point out that hydrogen is not completely 
ionized in the radio-emitting region and that this must be
taken into account when estimating the mass-loss rate
(Wright and Barlow 1975). In order to fit H and He lines simultaneously
it was necessary to use a slow velocity law ($\beta$=2.5) and
a photospheric velocity of $\sim$30$~{\rm km}~{\rm s}^{-1}$.  
It appears that P Cygni's absorption-line profiles are very sensitive to
small changes in the mass-loss rate. The latter alters the ionization
structure and affects sub-millimeter and radio fluxes. It is
quite possible that the variations in radio flux observed by
van den Oord et al. (1985) were due to such changes in
ionization structure. 

Detailed modelling of P Cygni's emission-line spectrum
and of the emergent continuum energy distribution for the 
wavelength range 2500 \AA\ $<$ $\lambda <$ 60$\mu$m
led Najarro et al. (1997) to the following values  
of the parameters describing the state of the 
atmosphere/wind and the position of the star on the H-R diagram:

\begin{description}
\item[]$R_{*}$=75\Rsun\
\item[]$L_{*}=5.6\times10^{5}$\Lsun\
\item[]\Teff=18\,200 K
\item[]n$_{\rm He}$/n$_{\rm H}$=0.3
\item[]\.M=3.0$\times10^{-5}$\Msun yr$^{-1}$
\item[]$V_{\infty}$=185~${\rm km}~{\rm s}^{-1}$
\end{description}

Lamers et al. (1983) derived $R_{*}$=76\Rsun\ from a study of 
the energy distribution of P Cygni (using ATLAS6 models) together 
with a distance of $\sim$1.8  kpc estimated from cluster membership. 
Such a distance is in agreement with the Hipparcos parallax of 
0.52$\pm$0.50 mas.

An important result from different spectroscopic analyses
of P Cygni is the enhanced helium abundance. The values
of n$_{\rm He}$/n$_{\rm H}$ proposed in the literature 
agree reasonable well; based on the H and He\,{\sc i} near-infrared lines 
Najarro et al (1997) and Deacon and Barlow (1991) 
suggest 0.4$\pm$0.15 and 0.5$\pm$0.1, respectively.

\section{Evolutionary status}

The high n$_{\rm He}$/n$_{\rm H}$ abundance ratio confirms 
that P Cygni has evolved away from the ZAMS. There are indications 
(Johnson et al. 1992) that P Cygni's nebula is composed of material 
in which most of the original carbon has been converted into
nitrogen by the CN cycle. Since it is known that the CN cycle
achieves equilibrium before the CNO cycle converts
all oxygen into nitrogen, the fact that we observe 
pseudo-photospheric absorption lines of oxygen indicating
a solar abundance of this element (Israelian 1995) suggests 
that the upper atmosphere has not yet been contaminated 
by the products of the CNO cycle.  
  
P Cygni is in a relatively quiescent phase of its 
post-main-sequence LBV life, exhibiting small-scale 
spectrophotometric variations but not undergoing the
moderate photometric outbursts observed in
S Dor, AG Car and other LBVs. The 1600 AD outburst
confirms that P Cygni must have crossed the Humphreys-Davidson
instability limit at least once. However, we do not know
whether it will suffer further outbursts in the future or how soon,
if ever, it will evolve towards a Wolf-Rayet-like phase.
Using the pulsational analysis of Kiriakidis et al. (1993),
Langer et al. (1994) proposed the following 
evolutionary sequence for very massive stars: 
O star $\rightarrow$ Of $\rightarrow$ H-rich WN star
$\rightarrow$ LBV $\rightarrow$ H-poor WN star  
$\rightarrow$ H-free WN star $\rightarrow$ WC $\rightarrow$ SN.
They have identified P Cygni with the hydrogen 
shell-burning phase and derived a current mass of 23$\pm$5\Msun.
This value of the mass, although in conflict with the 50\Msun\ 
suggested by El Eid and Hartmann (1993), does fit 
stellar parameters of the star obtained from a detailed
spectroscopic analysis by Najarro et al. (1997). 

The mechanism responsible for the giant and moderate variations 
of LBVs has been intuitively linked to a pulsation
instability. The understading of this mechanism 
is of great importance and will help to 
understand the intermediate phases in the evolutionary
chain O-star $\rightarrow$ LBV $\rightarrow$ WR-star. 
Maeder (1989) proposed density inversions that would cause strong 
instabilities in the atmospheres of LBVs. Inverted density gradients 
may lead to pulsation solutions (so called strange modes) that can
grow rapidly in amplitude. These pulsations lie in regions where 
the supra-Eddington luminosity causes an opacity peak. The rapid
growth of the density inversion results in an outburst.
The star moves again to the blue (away from the HD limit) when the
luminosity decreases to below the Eddington limit in the now exposed 
deeper layers and the opacity becomes so low that it can no longer 
drive the catastrophic mass loss. A high helium abundance tends
to stabilize the outburst by decreasing the pulsation amplitude. 
Stothers and Chin (1993, 1995) have
investigated a dynamical instability of massive stars 
in the linear adiabatic approximation and found a relation
between its occurrence and values below 4/3 of the mean adiabatic
exponent $\Gamma_{1}$ averaged over a conveniently chosen 
part of the stellar envelope. Their results have not been
confirmed by Glatzel and Kirakidis (1998) who also questioned
the validity of the adiabatic approximation for a stability 
analysis of massive stars. However, recently Stothers (1999a)
has performed a numerical stability analysis based on linear
and nonlinear hydrodynamical models of nonadiabatic, spherically
symmetric stellar envelopes, and confirmed results of
Stothers and Chin (1993, 1995) that the purely adiabatic 
criterion $\Gamma_{1} > 4/3$ does in fact determine dynamical
stability. It has been also shown (Stothers 1999b) that  
rotation is not  affecting directly the possible course of stellar 
evolution. For LBVs, evolving in a late stage of helium-core
burning, the luminosity-to-mass ratio would probably be increased
by rotation.  
 
The dependence of nonadiabatic instabilities on stellar
parameters is linked with the properties of the corresponding
opacity maxima (Fe, He or He/H opacity bumps) and their
positions within the star. Numerical simulations (Cox et al.
1997; Guzik et al. 1997) of the evolution of strange modes 
show that the outburst occurs when relatively deep atmospheric
layers exceed the Eddington luminosity causing a sudden
increase of the photospheric radial velocity (up
to 200~${\rm km}~{\rm s}^{-1}$). The rapid opacity rise during 
a pulsation cycle will lead to supra-Eddington luminosities.

The question is whether in a real star the convection will turn
rapidly enough to transport the radiation. To be more
realistic, these numerical simulations should consider at least 
1) time-dependent convection, 2) a large number of grid points,
and 3) to follow the material as it leaves the stellar surface.
The last point is added in order to bring the theory closer
to the observations. It is also not clear whether
these models are predicting (or aiming to predict)
moderate or giant eruptions. In general, the results of
such simulations must be interpreted with caution.

\section{\bf Some conclusions}

Even after a study of 400 years, at times particularly intensive during
this century, P Cygni continues to baffle us. Surely enough, progress has
been made, many observations using different techniques and covering almost
all wavelength regions have been made, various aspects of the star's 
behaviour have come to be known, and some of the latter have begun to
be understood. Despite all this, the number of unanswered questions
will be capable of occupying the investigative efforts of a whole generation
of researchers. Without pretending to be complete, here we offer a list,
in no particular order, of the more important questions deserving our
attention in our 21st and P Cygni's 5th century.

\begin{description}
\item[]P Cygni has shown us that to disentangle the combined effects of a
   variety of physical processes, the value of long series of observations
   cannot be overestimated. This approach needs to be continued and also 
applied
   more frequently to other objects.
\item[]The mechanism causing giant outbursts in LBVs is not yet understood. One
   senses that the mass of the star is the most important parameter, but
   how differences in mass are manifested in different outbursts and in the
   overall history of the star, is still a matter of contention that requires
   both more observations and more calculations. In this respect, the apparent
   low frequency of giant outbursts seems to indicate that we will need some 
   patience before getting to the bottom of this question.
\item[]While the short-term variations, both in brightness and in spectral-line
   profiles, are there to be observed all the time, their mechanism is still
   surrounded by a nebulosity of unanswered questions. The influences of
   rotation, pulsations, and magnetic fields, just to name a few, are not yet
   sufficiently understood.
\item[]We chose the title ``P Cygni: An Extraordinary LBV'' for this review.
   Is the adjective `extraordinary' due to the fact that P Cygni, despite
   being the first of something, is not a good example of a LBV? Whatever the
   answer to this question, does P Cygni teach us valuable lessons about LBVs
   in general, or are all such stars peculiar in one way or other and need
   an individual approach? Do small differences in physical parameters
   produce vastly different phenomena in the upper regions of the HR diagram?
\end{description}

It is the authors's hope that this review may be of some help to keep the
interest in P Cygni alive so that further progress towards a more complete
understanding of this `extraordinary LBV' can be made.

%\begin{acknowledgements}
\section*{Acknowledgements}
The authors are grateful to Dr. N.~Markova  for stimulating discussions
and to the referees R.~M. Humphreys, O.~Stahl, and another
unknown one for their constructive remarks. This work made use of the 
STARLINK network. MdG thanks DENI and PPARC for support.
%\end{acknowledgements}

\end{article}

\clearpage
\def\baselinestretch{1}

\begin{table*}
\caption{Radial-velocity data (in ${\rm km}~{\rm s}^{-1}$) of DACs 
in P Cygni reported in the literature.}
\begin{center}
\begin{tabular}{lllll}
\hline
ion & variable DAC &  IP(eV) & EP(eV) & References \\
\hline

Balmer   & 90 to 230  &      &  	  & de Groot 1969, Ivanova et al.1982, 
Markova
1986a\\
HeI      & 90 to 210  & 0.00 & 20.5 to 21.1 & Herbig 1962, Herman 1964, 
Astafyev 1968\\
         &            &      &  	  & de Groot~1969,  Ivanova et al. 1982, 
Markova~1993b\\
NaI      & 185        & 0.00 & 0.00	  & Ozemre 1978 \\
         & 188        &      &  	  & Markova 1991 \\
CaII (UV)& 128        & 6.09 & 0.00	  & Luud and Sapar 1980 \\
CaII     & 90 to 200  &      & 0.00	  & Viotti and Nesci 1973, Markova 1990 \\
         & 170 to 200 &      &  	  & Luud, Golandski and Iarigina~1975, Ozemre 
1978 \\
         & 130        &      &  	  & Beals 1950 \\
         & 185        &      &  	  & Ozemre 1978 \\
CrII (UV)& 177        & 6.74 & 1.5	  & Cassatella et al. 1979, Lamers et al. 
1985 \\
         & 170        &      &  	  & Luud and Sapar 1980 \\ 
MnII (UV)& 100 to 180 & 7.40 & 0.0 to 5.4 & Cassatella et al. 1979, Lamers et 
al. 1985 \\
         & 153        &      &  	  & Luud and Sapar 1980 \\
NiII (UV)& 100 to 180 & 7.61 & 0.0 to 6.8 & Cassattella et al. 1979,Lamers et 
al. 1985 \\
         & 182        &      &  	  & Luud and Sapar 1980 \\ 
MgII (UV)& 158        & 7.61 & 0.0 to 4.4 & Luud and Sapar 1980 \\
MgII     & 70 to 180  & 7.61 & 8.83       & Markova 1991 \\
         & 95 to 124  &      &  	  & Ivanova et al. 1982 \\
FeII (UV)& 112 to 174 & 7.86 & 0.9 to 2.6 & Cassatella et al. 1979, Lamers et 
al. 1985 \\
         &  90 to 180 &      &  	  & Israelian et al. 1996 \\
     (UV)&  180       &      &  	  & Luud and Sapar 1980 \\
FeIII(UV)& 100 to 157 &16.16 & 7 to 12    & Lamers et al. 1985 \\
     (UV)&  60 to 170 &      &  	  & Israelian et al. 1996 \\
     (UV)& 133        &      &  	  & Luud and Sapar 1980 \\
     (UV)& 157        &      &  	  & Hutchings 1979 \\
FeIII    & 90 to 180  &      & 8.21	  & Herbig 1962, Markova 1998\\
SII  (UV)& 178        &10.31 & 0.00	  & Luud and Sapar 1980 \\
PII  (UV)& 163        &10.90 & 0.00	  & Luud and Sapar 1980 \\ 
OII      & 50 to  90  &13.56 & 22.9  	  & Markova 1993a \\
TiIII(UV)&112         &13.6  & 0.0 to 4.7 & Cassatela et al. 1979 \\
     (UV)&137         &      &  	  & Luud and Sapar 1980 \\
NII      & 90 to 130  &14.49 &18.4        & Markova 1993a \\
MnIII(UV)& 89         &15.64 &  	  & Luud and Sapar 1980 \\ 
SiIII(UV)& 71         &16.27 &1.0 to 17.7 & Cassatella et al. 1979 \\
     (UV)& 110        &      &  	  & Luud and Sapar 1980 \\
     (UV)& 140        &      &  	  & Hutchings 1979 \\
SiIII    & 70 to 115  &      &18.92	  & Markova 1993a \\
CrIII(UV)& 92         &16.6  & 2.1 to 8.9 & Cassatella et al.\\
     (UV)& 123        &      &  	  & Luud and Sapar 1980 \\
NiIII(UV)& 79         &18.4  &  	  & Cassatella et al. 1979 \\
     (UV)& 131        &      &  	  & Luud and Sapar 1980 \\
AlIII(UV)& 202        &18.75 & 0.00	  & Luud and Sapar 1980 \\
AlIII(UV)& 159        &      & 6.6	  & Luud and Sapar 1980 \\
SiIV     & 20 to 70   &33.32 & 23.95	  & Markova 1993a \\	     
         & 40 to 80   &      &  	  & Ivanova et al. 1982 \\
\hline
\end{tabular}
\end{center}
\end{table*}

\clearpage
\begin{itemize}

\item Table 1.
Radial-velocity data (in ${\rm km}~{\rm s}^{-1}$) of DACs
in P Cygni reported in the literature.

\begin{figure}
\centerline{\psfig{file=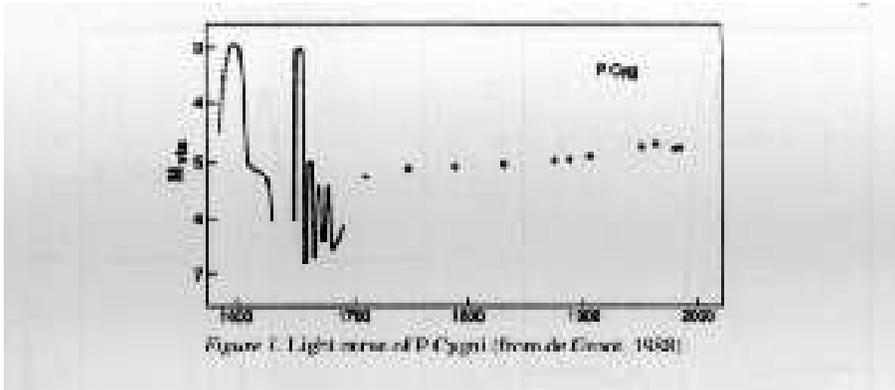,width=12cm}}
\caption[]{Light curve of P Cygni (from de Groot 1988)
}
\end{figure}

\item Fig 1.
Light curve of P Cygni (from de Groot 1988).

\begin{figure}
\centerline{\psfig{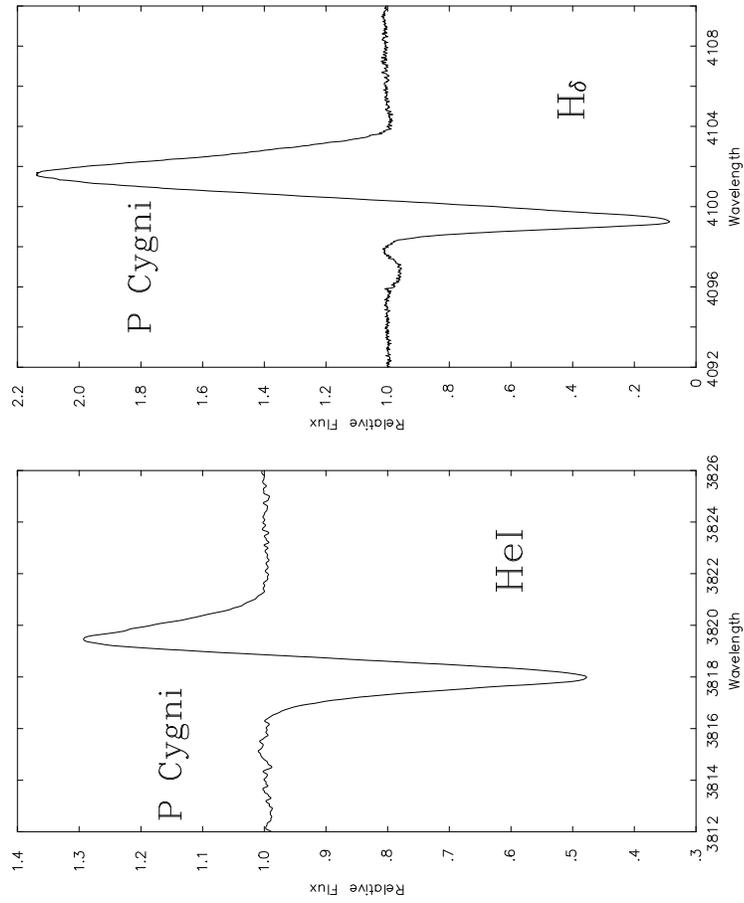}}
\caption[]{Typical P Cygni-type profiles of
HeI (3819.6 \AA) and H$\delta$ lines.
The spectrum was  obtained 1999 May 29 with the 2.5-m Nordic Optical
Telescope of the La Palma Observatory (Canary Islands, Spain)
by G. Israelian. The spectrum has a resolution 80\,000 and a
signal-to-noise (S/N) $\sim$ 200.}
\end{figure}

\item Fig 2.
Typical P Cygni-type profiles of HeI (3819.6 \AA) and H$\delta$ lines.
The spectrum was  obtained 1999 May 29 with the 2.5-m Nordic Optical
Telescope of the La Palma Observatory (Canary Islands, Spain)
by G. Israelian. The spectrum has a resolution 80\,000 and a
signal-to-noise (S/N) $\sim$ 200.

\begin{figure}
\centerline{\psfig{file=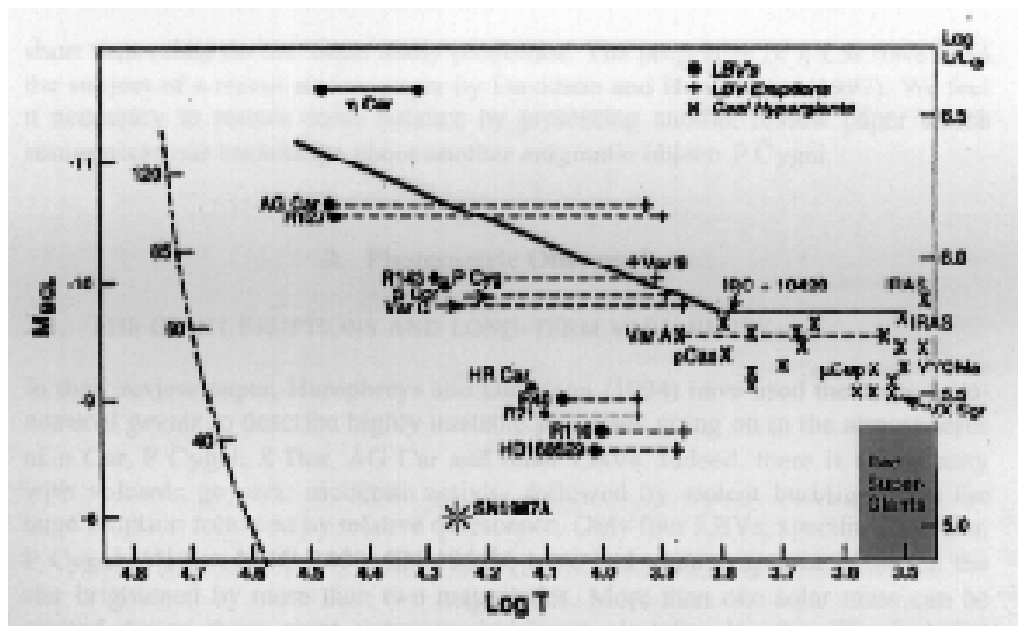,width=12cm}}
\caption[]{The most luminous stars on the HR diagram. The dashed lines are
the LBV transitions from the quiescence (filled circle) and to the 
eruption phase (crosses). The Humphreys-Davidson upper luminosity 
limit is shown as a solid line (adopted from Humphreys and Davidson 1994).}
\end{figure}

\item Fig 3.
The most luminous stars on the HR diagram. The dashed lines are
the LBV transitions from the quiescence (filled circle) and to the 
eruption phase (crosses). The Humphreys-Davidson upper luminosity 
limit is shown as a solid line (adopted from Humphreys and Davidson 1994).

\begin{figure}
\centerline{\psfig{file=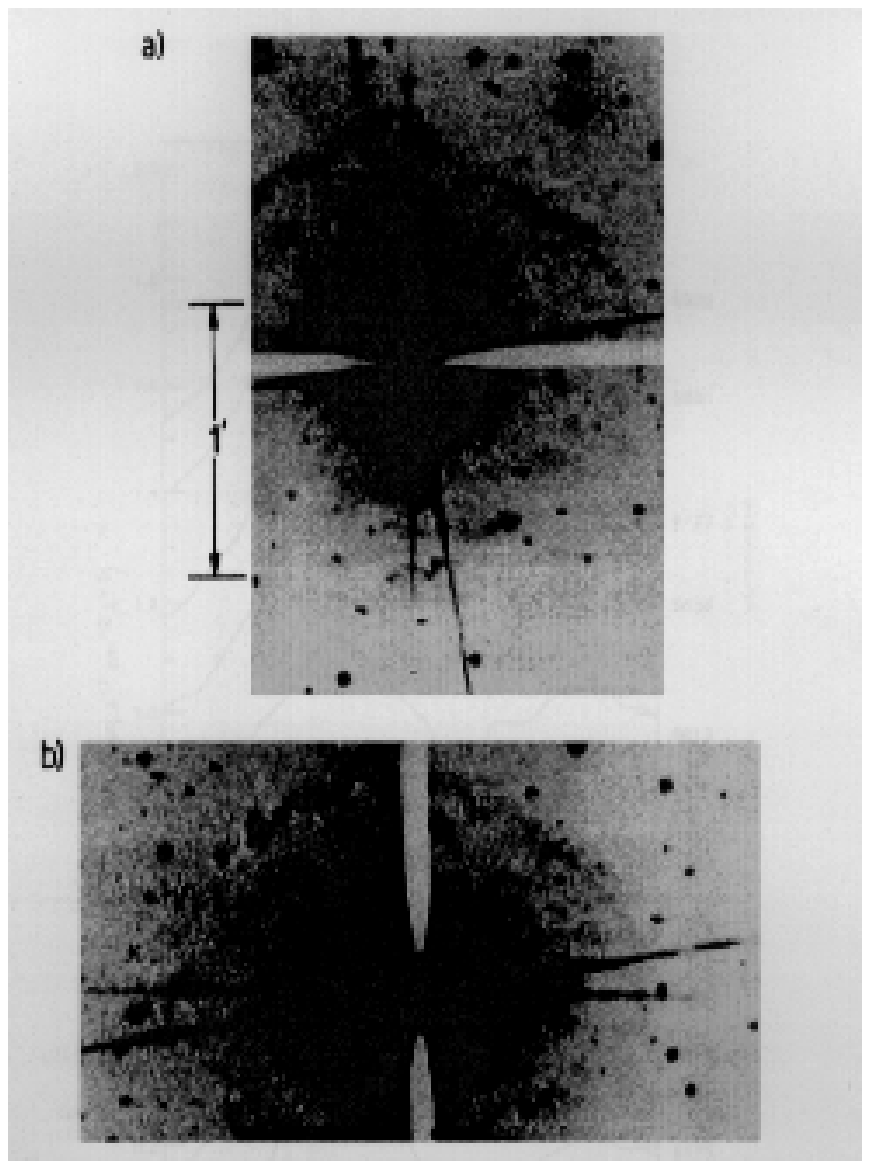,width=12cm}}
\caption[]{Negative grey-scale images of the outer nebulosity in the [NII]
6548 \AA\ line. The images were obtained at the La Palma Observatory
by Barlow et al. (1994) using the 4.3 arcsec wide occulting strip
oriented east-west and north-south, respectively.}
\end{figure}

\item Fig 4.
Negative grey-scale images of the outer nebulosity in the [NII]
6548 \AA\ line. The images were obtained at the La Palma Observatory
by Barlow et al. (1994) using the 4.3 arcsec wide occulting strip
oriented east-west and north-south, respectively.

\begin{figure}
\centerline{\psfig{file=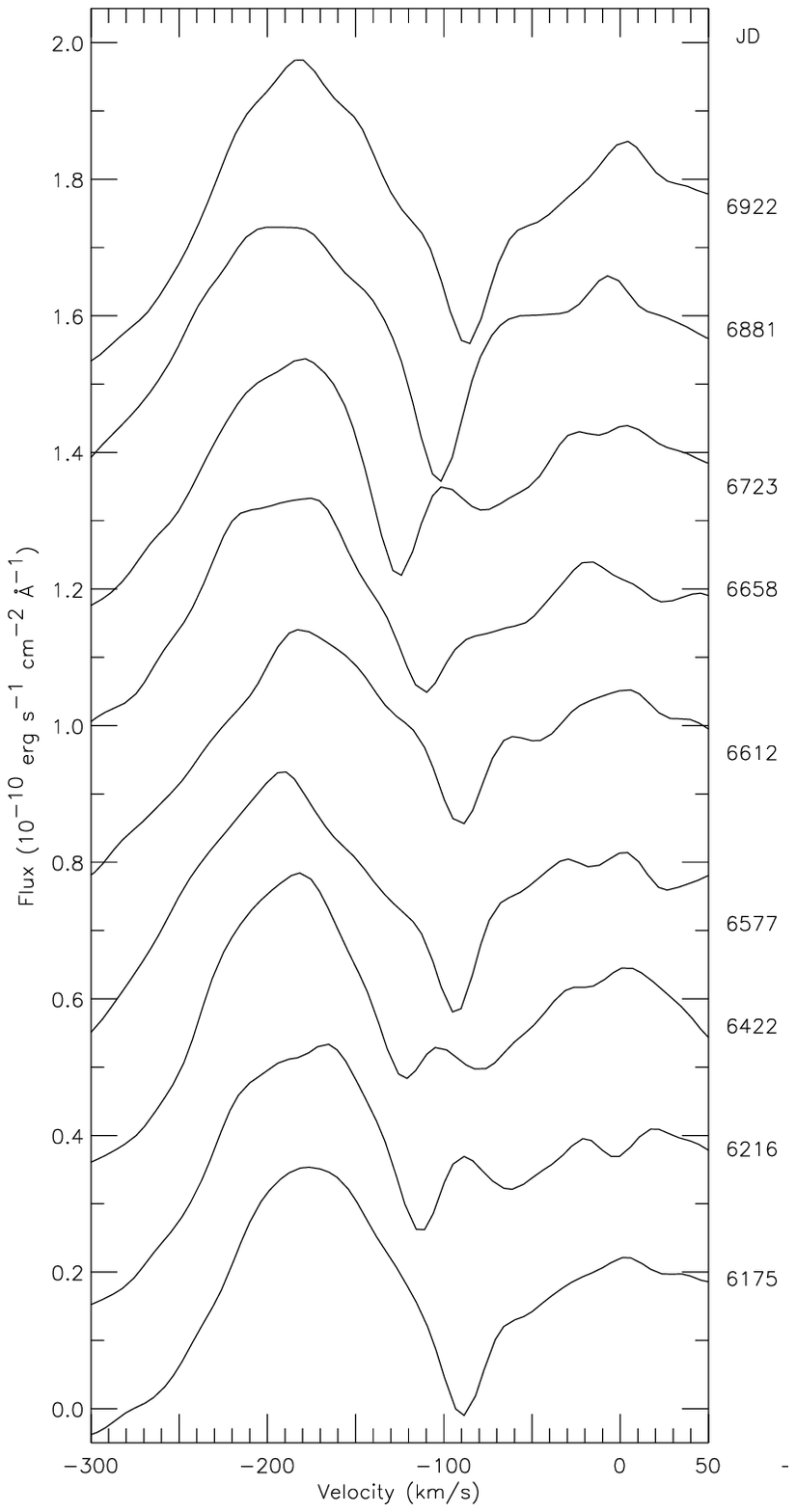,width=12cm}}
\caption[]{Profiles of the 1873 \AA\ line in the series of IUE spectra
SWP25731-30923 (JD6175-6922) (from Israelian et al. 1996).}
\end{figure}

\item Fig 5.
Profiles of the 1873 \AA\ line in the series of IUE spectra
SWP25731-30923 (JD6175-6922) (from Israelian et al. 1996).

\begin{figure}
\centerline{\psfig{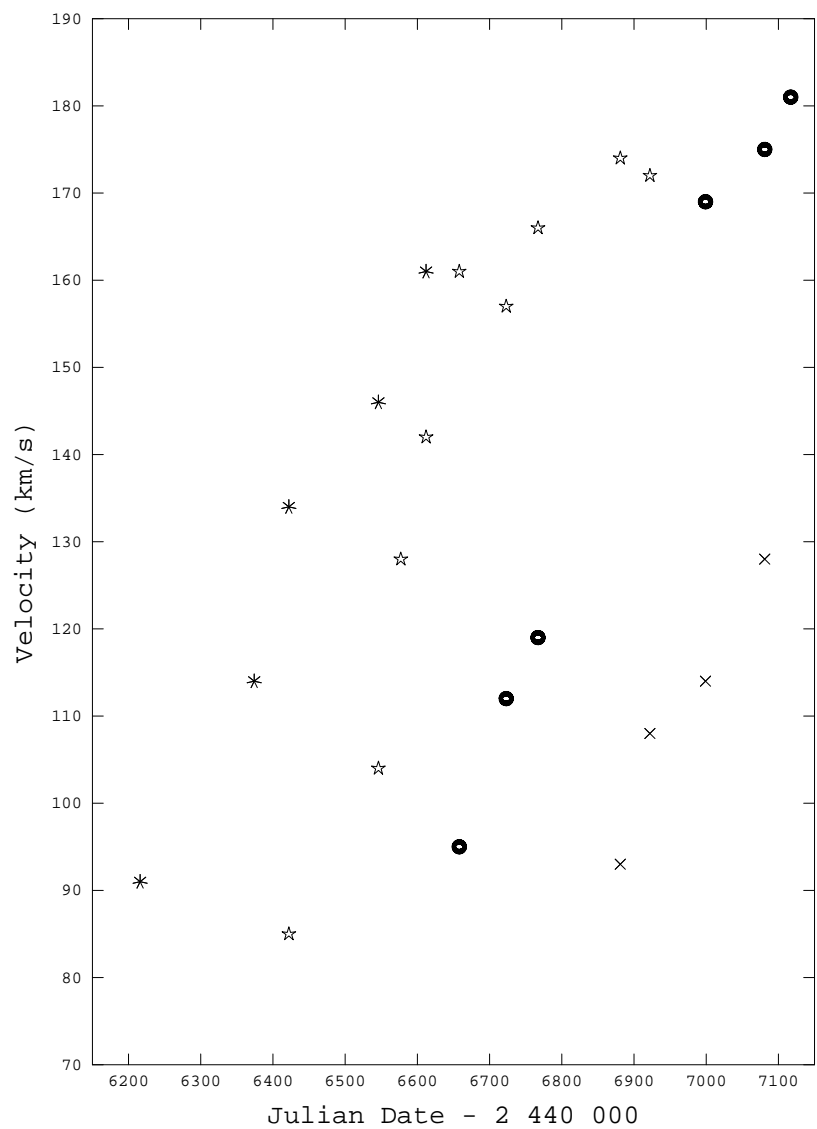}}
\caption[]{Variations of mean radial velocities of four shells
identified in FeII lines (from Israelian et al. 1996).
}
\end{figure}

\item Fig 6.
Variations of mean radial velocities of four shells
identified in FeII lines (from Israelian et al. 1996).

\end{itemize}

\end{document}